\documentclass[twocolumn,aps]{revtex4}
\usepackage{graphicx}% Include figure files
\usepackage{dcolumn}% Align table columns on decimal point
\usepackage{bm}% bold math
\usepackage[bf]{subfigure}
\usepackage{rotating}
\usepackage{helvet}
\usepackage{amssymb}

\begin{document}
%
%-----------------------------------------------------------------------
\title{Sensitivity of complex networks measurements}

%-----------------------------------------------------------------------
\author{P.\ R.\ Villas Boas}
\author{F.\ A.\ Rodrigues}
\author{G.\ Travieso}
\author{L.\ da\ F.\ Costa}
%-----------------------------------------------------------------------
\affiliation{Instituto de F\'{\i}sica de S\~{a}o Carlos,
Universidade de S\~{a}o Paulo, PO Box 369, 13560-970, S\~{a}o
Carlos, SP, Brazil }
%-----------------------------------------------------------------------

\begin{abstract}
Complex networks obtained from the real-world networks are often
characterized by incompleteness and noise, consequences of limited
sampling as well as artifacts in the acquisition process.  Because the
characterization, analysis and modeling of complex systems underlain
by complex networks are critically affected by the quality of the
respective initial structures, it becomes imperative to devise
methodologies for identifying and quantifying the effect of such
sampling problems on the characterization of complex networks.  Given
that several measurements need to be applied in order to achieve a
comprehensive characterization of complex networks, it is important to
investigate the effect of incompleteness and noise on such
quantifications. In this article we report such a study, involving 8
different measurements applied on 6 different complex networks models.
We evaluate the sensitiveness of the measurements to perturbations in
the topology of the network considering the relative entropy.  Three
particularly important types of progressive perturbations to the
network are considered: edge suppression, addition and rewiring.  The
conclusions have important practical consequences including the fact
that scale-free structures are more robust to perturbations.  The
measurements allowing the best balance of stability (smaller
sensitivity to perturbations) and discriminability (separation between
different network topologies) were also identified.
\end{abstract}

% PACS, the Physics and Astronomy % Classification Scheme.
% 89.75.Fb Structures and organization in complex systems
% 02.10.Ox Combinatorics; graph theory

%>>>>>>>>>>>>>>>>>>>>>>>>>>>>>>>>>

%-----------------------------------------------------------------------
\maketitle

\section{Introduction}

Complex networks theory has been largely applied to model real-world
systems, such as the Internet, the World Wide Web, protein
interactions, airlines, roads, food webs and
society~\cite{Boccaletti06:PR,Costa08:survey,Costa08:Surveyapp}. The
success of this area is to a great extent the consequence of two
recent developments: increase of computational power and availability
of several databases. In the former case, computers allowed processing
of networks with thousand or even million of vertices. In the latter,
many maps of interactions, raging from biology to social science, have
become available since the 90's. However, most of these maps are not
complete and methods should be developed to characterize these
networks~\cite{Stumpf05:PNAS}.

Sampling is a fundamental problem in complex networks because the
connectivity of many studied real-world networks may differ
substantially from the original complex systems from which they were
derived. This effect results in biased models, inaccurate
characterization, or incorrect classification and modeling of complex
systems.  In addition, many dynamical processes such as resilience to
random and target attacks~\cite{Albert:2000}, spreading
process~\cite{Boguna03:PRL}, synchronization~\cite{Boccaletti06:PR},
random walk~\cite{Costa07:PRE} and flow~\cite{Tadic07:IJBC} are
closely related to the completeness of networks.

A variety of sampling methods can be considered to map a complex
system into a network.  The sampling issue has been recently
considered in the analysis of different cross-section approaches to
construct biological, information, technological and social
networks. For instance, the available protein-protein interactions
cover only a fraction of the complete interactome map.  As a matter of
fact, the high-throughput ``yeast two-hybrid'' assay tends to provide
a high number of false positives, i.e.\ interactions identified in the
experiment but that never take place in the
cell~\cite{Mrowka01,Saito02:NAR}. Sprinzak et
at.~\cite{Sprinzak03:JMB} suggested that the reliability of the
high-throughput Y2H is about 50\%. Generally, it is assumed that the
incomplete maps can be extrapolated to the complete interactome, so
that limited sampling would not affect the topological structure of
the network~\cite{Jeong01:Nature}. This assumption is based on the
scale-free structure of protein interaction networks. However, the
subnetworks obtained by sampling of scale-free networks are not
guaranteed to be scale-free~\cite{Stumpf05:PNAS}. In addition, limited
sampling can result in scale-free structures irrespective of the
original network topology~\cite{Han05:NB, Khanin06:JCB}. In order to
overcome these limitations, efforts have been developed to obtain more
accurate databases of protein interactions~\cite{Krogan06:Nature}.

In the case of the World Wide Web, the network structure depend
strongly on the web crawler applied for sampling each chosen
domain~\cite{Serrano07}. Different sampling strategies can induce
bias, affecting in many ways the resulting recovered
structure~\cite{Becchetti06:PWLA}. Indeed, some crawlers tend to
overestimate the average number of connections of pages. A possible
solution for such limitations is to start from as large a set of pages
as possible~\cite{Becchetti06:PWLA}.

Accurate topologies of the Internet are fundamental for routing
strategies and to forecast its growth.  Internet sampling is generally
based on \emph{tracerouters} --- packets are sent through the network
in order to obtain the IP address of the routers in the path. However,
it is often assumed that these packets follow the shortest paths in
the network~\cite{Leguay07:CN}, implying a large set of connections to
be missed because of the possible presence of redundant links among
routers. Moreover, in the traceroute strategy edges close to the root
are more visible, i.e.\ the probability to obtain a edge far from the
root decreases with the distance from the
root~\cite{Lakhina03:INFOCOM}. It has also been observed that the
traceroute sampling of random networks leads to networks with
power-law degree distribution~\cite{Lakhina03:INFOCOM}.

Social networks are also incomplete. Generally, these networks are
restricted to a special class of human activity (e.g.\ music, sports,
casting and collaborations in science) or are constructed by
considering human relations (e.g.\ friendship and relationship). The way
in which these networks are obtained can often result in biased data,
such as the boundary specification problem, inaccuracy in
questionnaire application and inaccessibility of
subjects~\cite{Kossinets06:SN}. Moreover, depending of the considered
type of personal relationship, it becomes particularly hard to define
the links. It is a difficult difficult to estimate the effects of
missing data in social networks.

In the light of the above discussion, it becomes clear that sampling
bias might induce properties not representative of the actual complex
networks, leading to incorrect characterization and modeling.  The
sampling problem can be tackled by considering the following three
possible approaches:
\begin{enumerate}
  \item Improvement of sampling methodologies,
  \item Development of methods to predict missing~\cite{clauset2008hsa}
      and wrong links,
  \item Determination of the most suitable measurements to characterize
      incomplete networks.
\end{enumerate}

The first strategy depends on the type of the network that one wants
to sample. The second involves assumptions about rules and constraints
for each network structure, such as hierarchical
organization~\cite{clauset2008hsa}. The third has the intrinsic
advantages of being applicable to all already existing networks as
well as providing the only alternative in cases where the sampling
problems cannot be completely avoided. Some strategies have also been
developed to minimize the incomplete sampling problem by applying
remedial techniques~\cite{Serrano07}.  The work reported in the
current article relates to the third of the strategies above, by
quantifying the influence of several types of perturbations on complex
networks measurements. The perturbation of the degree distribution has
been investigated before~\cite{petermann2004esf, Han05:NB}.
Nevertheless, it is now realized that a single type of measurement
(i.e. degree) is not enough to characterize the structure of
networks~\cite{Costa:survey}. In addition, Alderson~\emph{et
  al.}~\cite{Alderson2005tts} have showed that networks with the same
degree distribution can present distinct topologies. Therefore, a
comprehensive set of measurement must be taken into account in order
to obtain an accurate network characterization~\cite{Costa:survey},
implying the effect of structural perturbations on several complex
networks measurements to become a particularly important issue.

Measurements that are too sensitive to perturbations in the network
may not be adequate to characterize incomplete or noisy
networks. Moreover, measurements that do not reflect differences
between distinct network structures are of reduced value because of
the implied lack of discriminability~\cite{Costa:survey}. In this
paper, we analyze the most traditional measurements used for networks
characterization by considering three important classes of
perturbations: (i) edge removal, (ii) edge inclusion, and (iii) edge
rewiring. Since these perturbations can be understood as noise added
to networks, we considered information processing
theory~\cite{cover1991eit} in order to quantify the sensitivity of
network measurements.  More specifically, we analyzed the distribution
of measurements in terms of relative entropies (Kullback-Leibler
distance). This measurement allows to determine the ``distance'' in
bits between two probability mass function. In this way, we obtained
the distribution of a given measurement $p$ and the distribution of
the same measurement after the network perturbation, $q$. The entropy
calculated taking into account these two distributions quantify how
much they are different (the relative entropy is always larger than
zero). Thus, by inspecting the behavior of the measurements under
these perturbations, we were able to identify the candidate
measurements most suitable for analysis and characterization of
networks constructed with incomplete data or in the presence of
noise. We analyzed 8 different measurements on 6 different complex
networks models.

\section{Basic concepts and methodology}

An undirected complex network (or graph) $G$ is defined as $G=(V,E)$,
where $V$ is the set of $N$ nodes and $E$ is the set of $M$ undirected
edges of the type $\{i,j\}$, indicating that the nodes $i$ and $j$ are
connected.  An undirected complex network without multiple edges can be represented in terms
of its adjacency matrix $A$, whose elements $a_{ij}$ and $a_{ji}$ are
equal to one whenever there is a connection between the vertices $i$
and $j$; and equal to 0 otherwise. Since most real-world networks are
composed by thousand or even million of vertices, the analysis of
their structure cannot be performed by visual inspection. In this way,
a set of measurements are considered in order to describe and
discriminate network topologies. These measurements can reflect
different features on the network, such as connectivity,
assortativity, centrality and hierarchies. In this work, we considered
the distribution of the following representative set of measurements
in order to characterize the network structures~\cite{Costa:survey}.
The chosen measurements include more traditional and simpler
measurements such as node degree and clustering coefficient as well as
more recent and sophisticate features such as betweeness centrality
and hierarchical measurements.

\begin{itemize}
\item \emph{Degree}: the degree of a node $i$, $k_i$, is given by its
  number of connections.

\item \emph{Average degree of nearest neighbors}: The average neighbor
  connectivity, $k_{nn}$, measures the average degree of the neighbors
  of the vertices in the network.

\item \emph{Clustering coefficient}: The clustering
  coefficient of a node $i$, $C_i$, is defined as the number of
  links between the vertices within its neighborhood, $l_i$, divided
  by the number of edges that could possibly exist between them
  ($k_i(k_i-1)/2$).

\item \emph{Hierarchical measurements}: Hierarchical measurements are
  defined by considering the successive neighborhoods around each
  node~\cite{Costa:2005a,Costa:survey}. Such measurements are
  particularly interesting because they reflect several topological
  scales around each reference node, from purely local (first
  neighborhood) to completely global (the most distance neighbors).
  The ring of vertices $R_d(i)$ (or hierchical/concentric level) is
  formed by those vertices distant $d$ edges from the reference vertex
  $i$.
  \begin{itemize}
    \item  \emph{Hierarchical degree} at level $d$, $hk_d(i)$, is
           defined as the number of edges connecting the rings
           $R_d(i)$ and $R_{d+1}(i)$.
    \item \emph{Hierarchical clustering coefficient} ,$hC_d$, is
           given by the number of edges among nodes in the
           respective $d$-ring ($m_{d}(i)$), divided by the total
           number of possible edges between the vertices in that
           ring.
    \item \emph{Divergence ratio}, $hdr_d$, corresponds to  the ratio
            between the number of vertices in the ring at level
            $d+1$ and the hierarchical node degree at level $d$.
  \end{itemize}

\item \emph{Shortest path length}: The shortest path length between
two vertices $i$ and $j$, $\ell_{ij}$, is given by the shortest
distance between that pair of vertices.

\item \emph{Betweenness centrality}: The betweenness centrality of a
vertex $i$, $B_i$, quantifies the fraction of shortest paths
between each pair of nodes in the network that pass through this
vertex.

\end{itemize}

\subsection{Perturbation methods}

In this work, the noise and incompleteness frequently found in complex
networks derived from real-world data are modeled in terms of three
basic types of structural perturbations, namely:
\begin{itemize}
\item Edge removal: Edges are selected at random and removed from the
  network.
\item Edge addition: Two not connected vertices are selected at
  random, and a connections is established between them.
\item Edge rewiring: Two pairs of connected vertices are chosen and
  their connections are interchanged.
\end{itemize}

In our analysis, we also considered a random combination of all these
types of perturbations.

The intensity of the perturbations ranged from 0 to 10\% of the total
number of edges in the network. In the case of the rewiring
perturbation, the number of steps necessary to reach 10\% of edges was
half of that required for the other two because each step corresponded
to a change of two edges. Perturbations involving vertices could also
be considered. However, the addition of vertices should depend on the
type of network in question. In order to make our analysis simpler and
more robust, we focused edge perturbations. The behavior of the
measurements was therefore studied with respect to several types of
edge perturbations.

\subsection{Relative entropy}

In statistical mechanics, the entropy is a measure of uncertainty or
disorganization in a physical system~\cite{reichl1980mcs}.  In
principle, the entropy is given by the logarithm of the number of ways
in which a system can be configured. The concept of entropy has many
application to different research areas. For instance, while in
quantum mechanics, the entropy is related to the von Neumann
entropy~\cite{vonneumann1932mgq}; in complexity theory, to the
Kolmogorov entropy~\cite{li1997ikc}. Here, we consider the concept of
entropy in the sense of information theory, where entropy is used to
quantify the minimum descriptive complexity of a random
variable~\cite{cover1991eit}. In this case, the entropy of a discrete
random distribution $p(x)$ is given as
\begin{equation}
H(p) = -\sum_{x}p(x) \log p(x),
\end{equation}
where the logarithm is taken on the base 2. In case of complex
networks, many of their properties result from heterogeneity of their
connections.  Jun \emph{et al.}~\cite{jun2007ner} suggested the
consideration of the normalized entropy of rank distribution in order
to analyze scale-free networks.

The \emph{relative entropy}, or Kullback-Leibler distance, measures
the ``distance'' in bits between two probability mass function $p(x)$
and $q(x)$ and is defined as
\begin{equation}
D(p,q) = \sum_x p(x)\log \frac{p(x)}{q(x)}.
\end{equation}\label{eq:relative_entropy}

Such value is always nonnegative and is zero if and only if $p =
q$. Typically $0\log\frac{0}{0} = 0$, $0\log
\frac{0}{q} = 0$ and $p\log\frac{p}{0} = \infty$.  Therefore, if $p$
is the distribution of a given network measurement and $q$ is the
distribution of the same measurement obtained from the respective
network under presence of noise, the relative entropy provides a sound
means to quantify the intensity of changes implied by the noise on the
distribution $p$. In this work we considered the relative entropy to
determine the sensitivity of different network measurements.

\subsection{Analyzed Networks}

In order to study the effects of perturbations on networks, we
considered structures generated by six different network models,
including traditional structures such as random, scale-free and
geographical models as well as more recent models such as limited and
non-linear preferential attachment.  The consideration of several models
is fundamental for investigating the effect of perturbations in
networks because the implied changes are strongly dependent on
specific network connectivity. Also, because of the distinct
properties of these models and that we can generate ensembles of networks, it becomes immediately possible to
quantify the discrimination of the measurements with respect to such
different types of structures.

\subsubsection{Theoretical models}

Since the perturbation dynamics can depend on the network structure,
we considered the following network models~\cite{Boccaletti06:PR}.
\begin{itemize}

\item \emph{Erd\H{o}s-R\'{e}nyi random graph (ER}): This model
  generates networks with random distribution of connections.  The
  network is constructed connecting each pair of vertices in the
  network with a fixed probability $p$~\cite{ErdosRenyi60:BISI}. This
  model generates a Poisson like degree distribution~\cite{Bollobas98}.

\item \emph{Small-world model of Watts and Strogatz (WS)}: To
  construct this type of small-word network, one starts with a regular
  ring lattice of $N$ vertices in which each vertex is connected to
  $\kappa$ nearest neighbors in each direction.  Each edge is then
  randomly rewired with probability $q$~\cite{Watts98:Nature}.

\item \emph{Barab\'{a}si-Albert scale-free model~(BA)}: This model
  generates networks with power law degree distribution.  The network
  is generated by starting with a set of $m_0$ vertices and, at each
  time step, the network grows with the addition of a new vertex
  with $m$ links. The vertices which receive the new edges are chosen
  following a linear preferential attachment rule, i.e.\ the
  probability of the new vertex $i$ to connect with an existing vertex
  $j$ is proportional to the degree of $j$, $\mathcal{P}(i\rightarrow
  j) = k_j/\sum_u k_u$~\cite{Barabasi99:Science}.

\item \emph{Waxman geographical model~(WG)}: Geographical networks can
  be constructed by distributing $N$ vertices at random in a 2D space
  and connecting them according to the
  distance~\cite{Waxman1988rmc}. This model is created by randomly
  distributing $N$ vertices in a square of length $L = \sqrt{N}$ and
  connecting them with probability $p = e^{-\lambda d}$, where $d$ is
  their geographic distance, and $\lambda$ is a constant adjusted to
  achieve the desirable average degree.

\item \emph{Limited scale-free model (LSF)}: The network is generated
  as in the BA model but the maximum degree is limited to a maximum
  $k_{max}$ value~\cite{amaral2000csw}.

\item \emph{Nonlinear preferential attachment network model~(NLBA)}: The network
  is constructed as in the BA model, but instead of a linear
  preferential attachment rule, the vertices are connected following a
  nonlinear preferential attachment rule, i.e., $P_{i\rightarrow j} =
  k_j^\alpha/\sum_u k_u^\alpha$. In this case, while for $\alpha < 1$,
  the network has a stretched exponential degree distribution, for
  $\alpha > 1$ a single site connects to nearly all other
  sites~\cite{krapivsky2001ogr}.
\end{itemize}

\section{Results and discussions}

Our simulations considered the following parameters: $N =1,\!000$
vertices; average degree 6; in case of WS model, the probability $q$
of reconnection was $0.3$; $\lambda$ was $1.0$ for WG model; $\alpha =
0.5$ for the NLBA network model; and the maximum degree was $k_{max} =
50$ for the LSF network mode.

The perturbations were performed from 0.5\% up to 10\% of the total
number of edges of each network in steps of 0.5\%. Also, for each
network model, 20 networks were generated at each step.  For each
network, we obtained the normalized distribution of measurements
considering 50 boxes. The histograms for every network were obtained
by taking into account the same maximum and minimum values for each
measurement. In order to quantify the variation on the distribution,
we calculated the relative entropies by considering the
equation~\ref{eq:relative_entropy}.

\begin{figure*}[!ht]
  \begin{center}
    \includegraphics[width=0.7\linewidth]{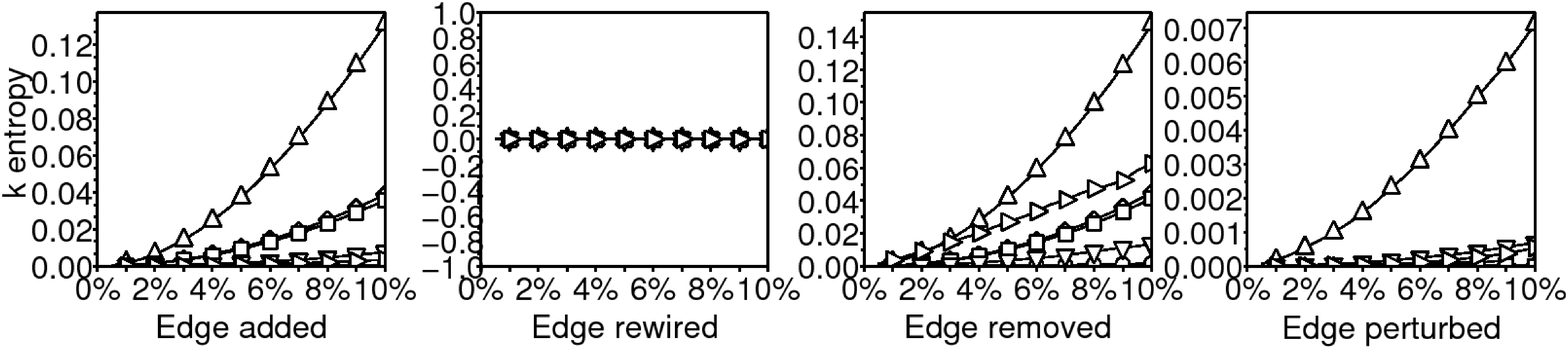}
    \includegraphics[width=0.7\linewidth]{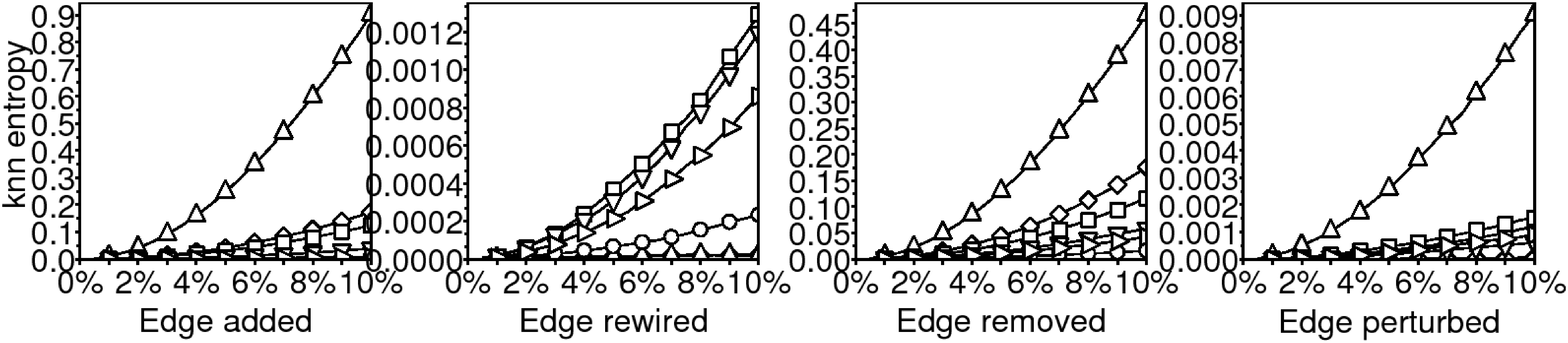}
    \includegraphics[width=0.7\linewidth]{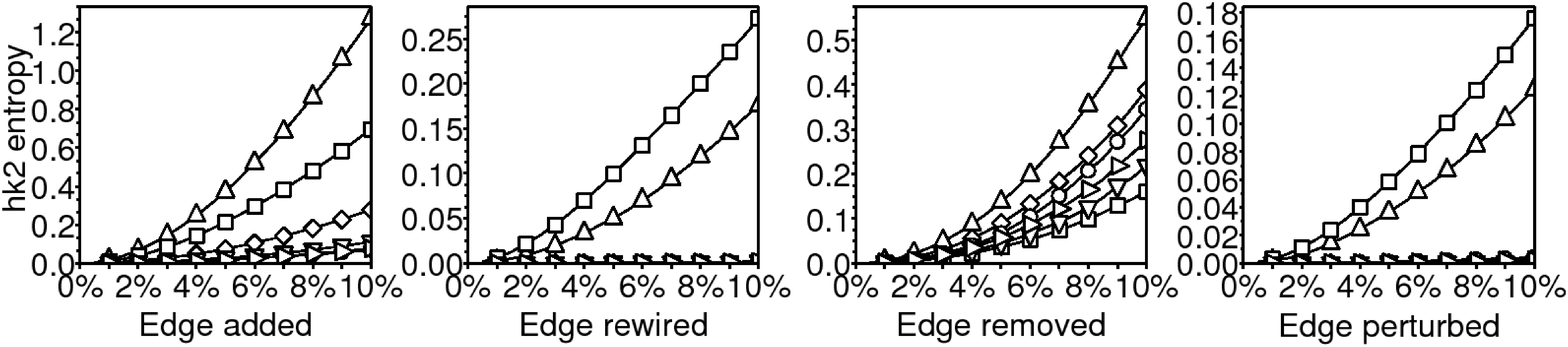}
    \includegraphics[width=0.7\linewidth]{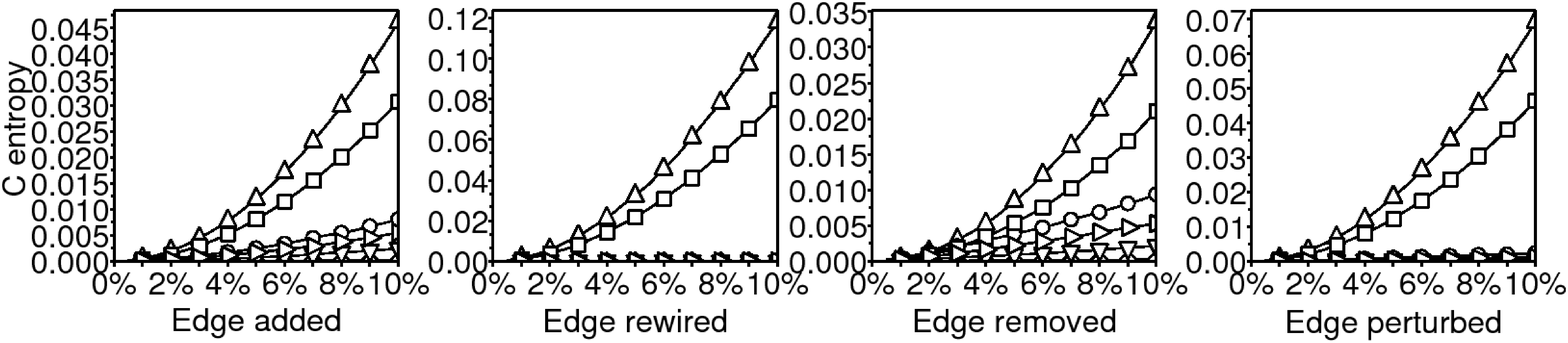}
    \includegraphics[width=0.7\linewidth]{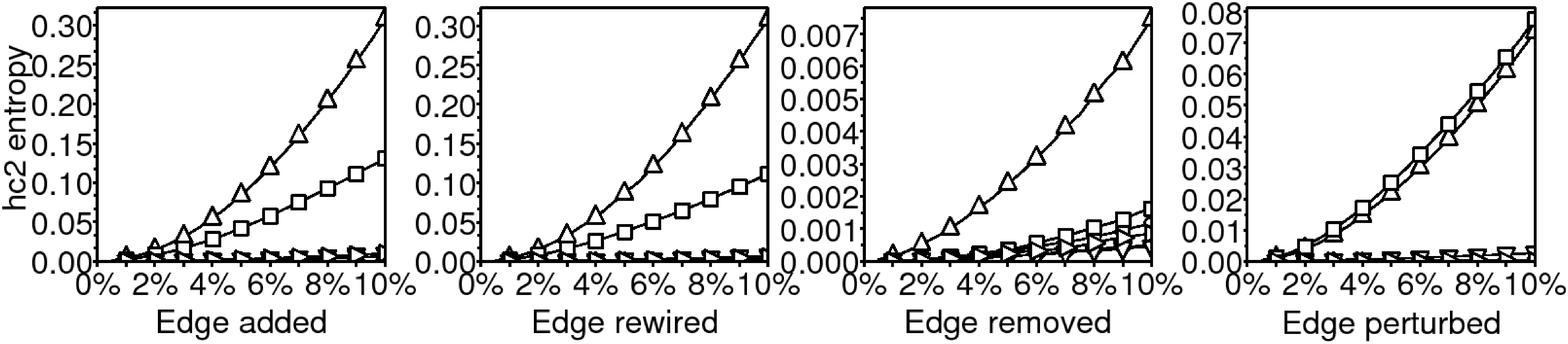}
    \includegraphics[width=0.7\linewidth]{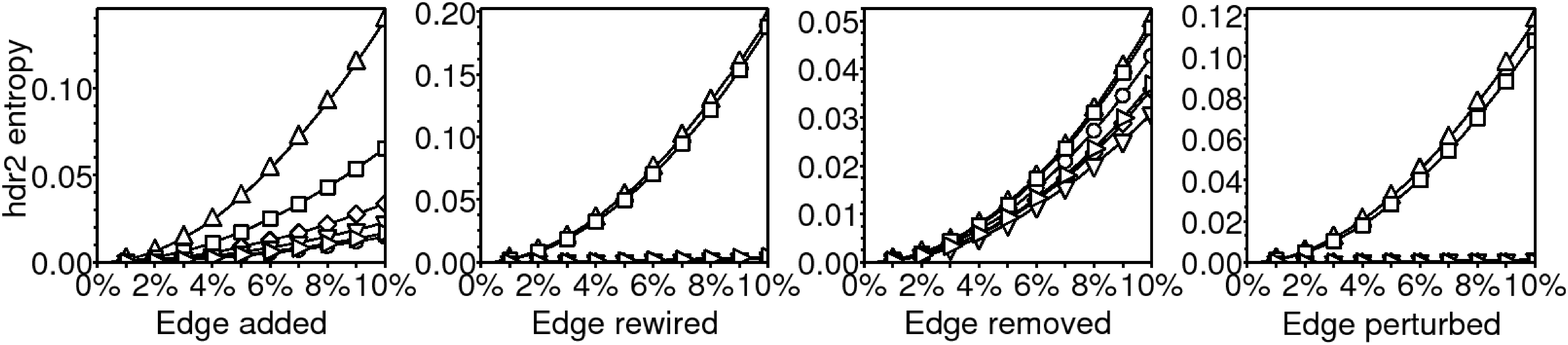}
    \includegraphics[width=0.7\linewidth]{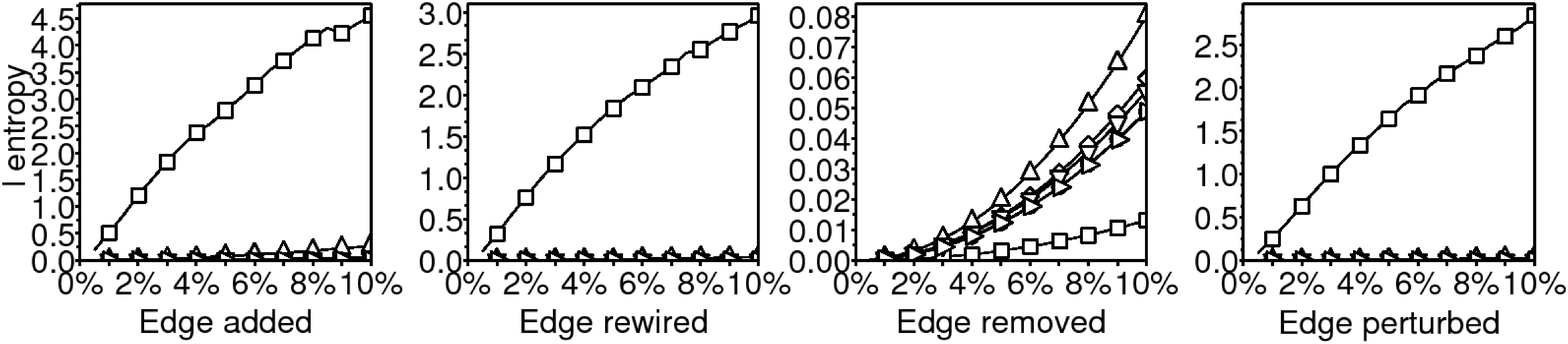}
    \includegraphics[width=0.7\linewidth]{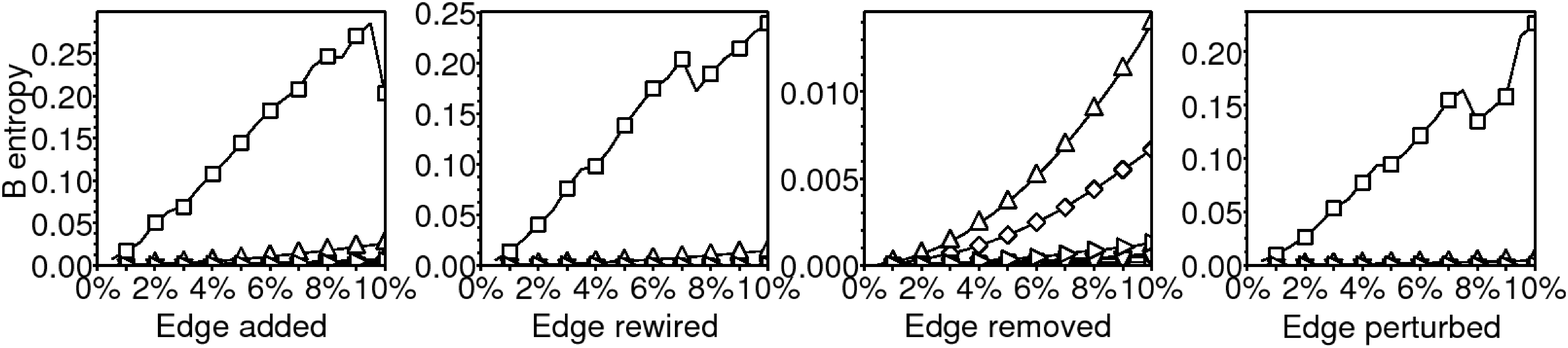}
  \end{center}
  \caption{Measurement entropies for the considered network models: \large
    $\lozenge$ \normalsize -- Erd\H{o}s and R\'enyi's random graph,
    $\triangle$ -- Watts and Strogatz's small-world model, \small
    {$\bigcirc$} \normalsize -- Barab\'asi and Albert scale-free
    model, \large $\square$ \normalsize -- Waxman's geographic model,
    $\bigtriangledown$ -- Krapvisky's non-liner preferential attachment model, and
    \Large$\triangleright$ \normalsize -- Amaral et al.'s limited scale-free model}
  \label{fig:pert_models}
\end{figure*}

Figure~\ref{fig:pert_models} present the results with respect to each of the 6 network models, and Table~\ref{Tab:Av_pert} shows the average of the relative entropy for all network models considering 10\% of edge perturbations. The main results observed are discussed as follows:
%??? Reescrevi o paragrafo abaixo?????
\begin{itemize}
\item Random edge removal causes smaller variations in the
  measurements than the other three types of perturbations (including
  the random combination of all perturbations). From this result, we
  can conclude that it is better not to include edges about which we are
  uncertain, as the inclusion of an unexistent edge implies larger
  deviations of the measurements than the removal of an existing one.

\item Comparing the results for all networks, according to each
  perturbation, the measurements can be ordered as following
  according to the values of the maximum entropy:
    \begin{itemize}
      \item Edge addition: \\$B$, $C$, $k$, $hdr_2$, $hC_2$,
                             $k_{nn}$,  $hk_2$, $\ell$.
      \item Edge rewiring:\\ $k$, $k_{nn}$, $C$, $hC_2$, $B$,
                             $hdr_2$, $hk_2$, $\ell$.
      \item Edge removing:\\ $hC_2$, $B$, $C$, $hdr_2$, $k$,
                             $\ell$, $k_{nn}$, $hk_2$.
      \item All three perturbation together: \\$k$, $hC_2$, $k_{nn}$,
                             $C$, $hdr_2$, $B$, $hk_2$, $\ell$.
    \end{itemize}
Depending on the perturbation, the measurements can be more or less sensitive. Therefore, it is important to select the appropriated set of measurements according to the type of perturbation, since particularly sensitive measurements can lead to wrong network characterization.

\item The shortest path length is the most sensitive network measurement,
  specially for the geographical WS and WG models. This result was expected
  because just a few rewiring in a regular network can lead to a
  small-world network~\cite{Watts98:Nature}.  In other words, adding
  or rewiring edges can connect vertices which are far away, reducing
  the average shortest path length. Therefore, this measurement is not
  particularly suitable for networks exhibiting geographical organization.

\item Among the network models, the scale-free structures resulted as those
  presenting the less sensitive structures, being the LSF and NLBA the most robust.
  Indeed, these network models generate topologies which are
  more close to real world networks than the BA model. For instance,
  the BA model tend to generate networks whose average clustering
  coefficient is smaller than that observed in real world networks. On
  the other hand, the LSF and NLBA can overcome this limitation by
  considering appropriated parameters.  The fact that scale-free
  networks are less sensitive to perturbation dynamics is a
  fundamental finding because most real-world networks are
  scale-free. Indeed, scale-free network have been previously observed
  to be highly resilience against random failures~\cite{Albert:2000},
  although just the average shortest paths length was investigated in
  that work. On the other hand, WG and WS models present the structures
  most sensitive to the considered perturbations.
\end{itemize}

\begin{table}
  \caption{Average variation of entropy for each plot in Figure~\ref{fig:pert_models} considering the maximum perturbation with respect to edges addition, edges rewiring, edge removing and all these perturbations applied together. }
\begin{tabular*}{0.48\textwidth}{@{\extracolsep{\fill}}l | c c c c c}
\hline
Meas.    &Addition   &Rewiring    & Removing   & Altogether\\
\hline
$k$              &0.0269   &0.0000   &0.0418   &0.0006 \\
$k_{nn}$    &0.1207    &0.0003   &0.0514    &0.0010 \\
$hk_2$       &0.3216    &0.0605    &0.2060   &0.0412 \\
$C$             &0.0127    &0.0197     &0.0050   &0.0122 \\
$hC_2$      &0.0354    &0.0245    &0.0003   &0.0008 \\
$hdr_2$    &0.0333    &0.0310     &0.0353    &0.0188  \\
$\ell$         &0.8218    &0.4992     &0.0510    &0.4771  \\
$B$            &0.0031    &0.0268     &0.0031    &0.0215 \\
\hline
Average           &0.1719	     &0.0828	 &	0.0493	&0.0717 \\
St. deviation   &0.2828		&0.1693       &0.0669	&0.1644 \\
\hline
\end{tabular*}\label{Tab:Av_pert}
\end{table}

In order to quantify the discriminability of each measurement, we
resourced again to the mutual information in order to obtain the
``distance'' between pairs of different types of models in the absence
of perturbations. So, high values of relative entropy suggest good
separability between models. Tables~\ref{Tab:degree} to~\ref{Tab:bet}
present the pairwise comparison for the six models for each of the eight measurements.
For instance, in the case of the degree distribution (see
table~\ref{Tab:degree}), the relative entropy between the BA and LSF
network models is the smallest one, followed by the relative entropy
between the ER and WG models. The betweenness centrality is the
measurement that provided the poorest separation between models. This
is mainly due to the lack of community structure of the considered
network models.

Analyzing the average values on the tables~\ref{Tab:degree}
to~\ref{Tab:bet}, the measurements that provide highest
discriminiation (highest values of entropy) are in order presented in
Table~\ref{Tab:av_disc}.  Interesting to note that the degree
distribution, which is largely considered to characterize complex
network models, does not performed particularly well in our
analysis. For instance, while the relative entropy between the ER and
BA degree distributions is 0.424, for the hierarchical clustering
coefficient of level 2 it is 5.581. Indeed, the hierarchical
measurements accounted for the best overall characterization of
network structures with respect to discriminability.  This property is
possibly related to fact that the hierarchical measurements take into
account a larger portion of the original network, therefore providing
a more comprehensive quantification of the local topology.

\begin{figure*}[ht]
  \begin{center}
    \includegraphics[width=1\linewidth]{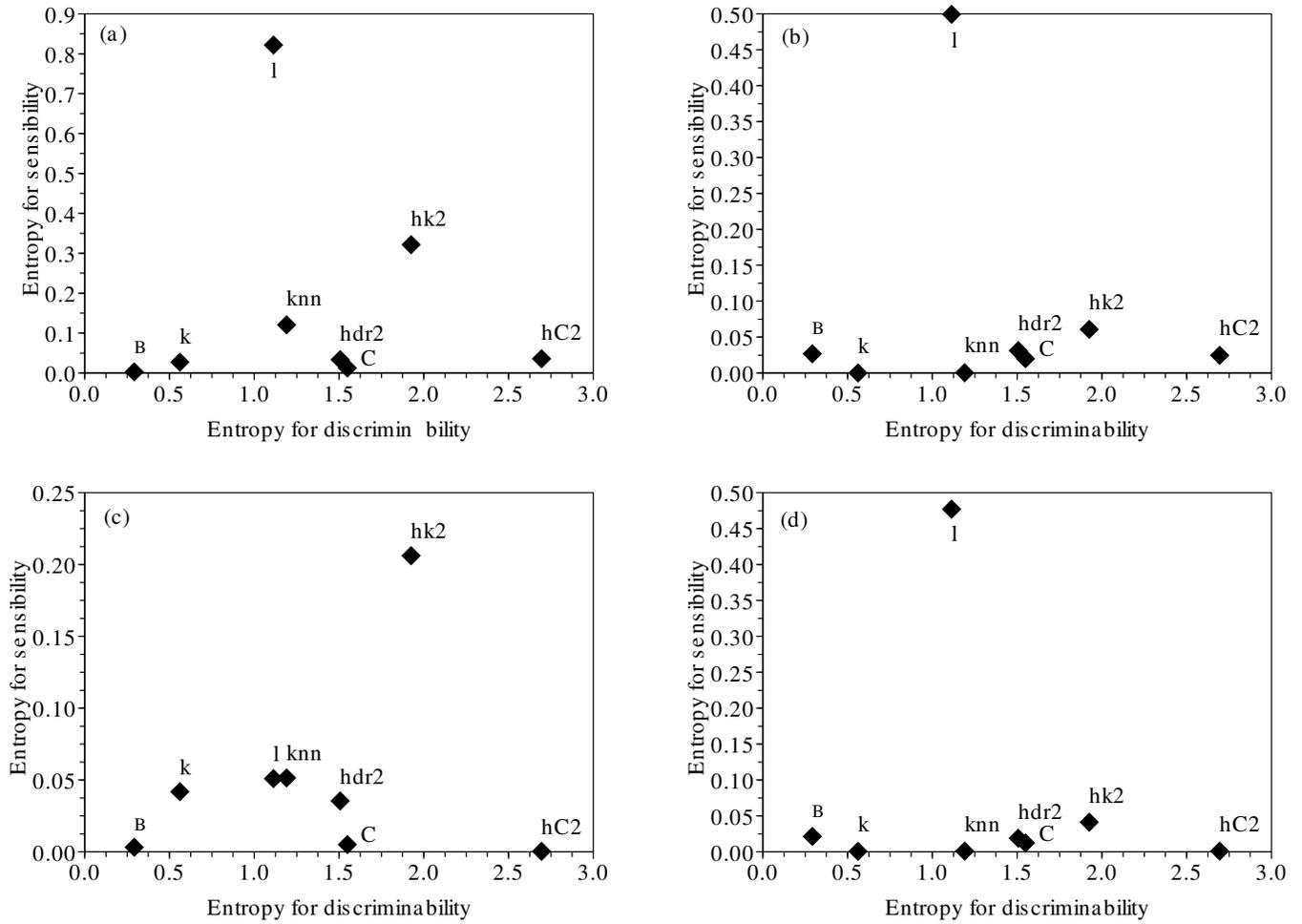}
  \end{center}
  \caption{Discrimination versus sensitivity of network measurements considering different types of perturbations, considering the values of the Tables~\ref{Tab:degree} to~\ref{Tab:bet} and Table~\ref{Tab:Av_pert}.}
  \label{fig:disc_sens}
\end{figure*}

The main motivation of our studying of perturbations in networks was
to find measurements allowing an acceptable compromise between
stability and discriminability. In this way, a proper measurement to
characterize sampled networks should be that which provide good
characterization of network structure and small sensitivity to
perturbations.  Figure~\ref{fig:disc_sens} shows scatterplots defined
by the sensitivity with respect to edge addition
(Fig.~\ref{fig:disc_sens}a), rewiring ((Fig.~\ref{fig:disc_sens}b),
removal (Fig.~\ref{fig:disc_sens}c) and joint perturbations
(Fig.~\ref{fig:disc_sens}d) versus the discriminability considering
the average relative entropy in each respective case.  The best
measurements are those resulting at the lower righ-hand portions of
these scatterplots.  Therefore, the measurements allowing the overall
best combinations of sensitivity and discriminability include the
$hC_2$, $hk_2$, $C$ and $hdr_2$.  Interestingly, the node degree,
betweeness and shortest path -- which have frequent and intensively
used for networks characterization -- are either too sensitive or not
discriminative.

\begin{table}
  \caption{Relative entropy of degree distribution, where ER is the
    Erd\H{o}s and R\'enyi random model, WS is the Watts and Strogatz
    small-world model, BA is the Barab\'asi and Albert scale-free
    model, WG is the Waxman geographic model, NLBA is the
    non-linear preferential attachment model, and LSF is the limited scale-free
    model.}
\begin{tabular*}{0.48\textwidth}{@{\extracolsep{\fill}}l | c c c c c c}
\hline
    & ER    & WS    & BA    & WG    & NLBA  & LSF   \\
\hline
 ER & 0.000 & 0.502 & 0.424 & 0.005 & 0.306 & 0.504 \\
 WS & 0.468 & 0.000 & 0.957 & 0.527 & 1.125 & 1.366 \\
 BA & 0.377 & 1.767 & 0.000 & 0.325 & 0.037 & 0.001 \\
 WG & 0.005 & 0.548 & 0.396 & 0.000 & 0.270 & 0.467 \\
 NLBA   & 0.527 & 1.708 & 0.047 & 0.544 & 0.000 & 0.040 \\
 LSF    & 0.704 & 2.056 & 0.022 & 0.719 & 0.082 & 0.000 \\
\hline
\end{tabular*}\label{Tab:degree}
\end{table}

\begin{table}
\caption{Relative entropy of average degree of nearest neighbors distribution.}
\begin{tabular*}{0.48\textwidth}{@{\extracolsep{\fill}}l | c c c c c c}
\hline
    & ER    & WS    & BA    & WG    & NLBA  & LSF   \\
\hline
 ER & 0.000 & 1.454 & 1.688 & 0.102 & 0.711 & 1.563 \\
 WS & 0.816 & 0.000 & 2.197 & 1.021 & 1.760 & 2.659 \\
 BA & 1.854 & 0.814 & 0.000 & 1.786 & 1.011 & 0.162 \\
 WG & 0.137 & 1.780 & 1.592 & 0.000 & 0.549 & 1.373 \\
 NLBA   & 1.312 & 1.677 & 0.627 & 1.372 & 0.000 & 0.387 \\
 LSF    & 1.907 & 0.375 & 0.121 & 2.291 & 0.638 & 0.000 \\
\hline
\end{tabular*}\label{Tab:knn}
\end{table}

\begin{table}
\caption{Relative entropy of hierarchical degree of level 2 distribution.}
\begin{tabular*}{0.48\textwidth}{@{\extracolsep{\fill}}l | c c c c c c}
\hline
    & ER    & WS    & BA    & WG    & NLBA  & LSF   \\
\hline
 ER & 0.000 & 2.621 & 2.272 & 2.386 & 0.562 & 1.597 \\
 WS & 1.949 & 0.000 & 6.353 & 0.276 & 3.158 & 5.447 \\
 BA & 1.353 & 0.374 & 0.000 & 1.271 & 0.741 & 0.088 \\
 WG & 1.655 & 0.428 & 6.319 & 0.000 & 3.212 & 5.406 \\
 NLBA   & 0.739 & 1.763 & 0.786 & 2.064 & 0.000 & 0.393 \\
 LSF    & 1.115 & 1.034 & 0.091 & 1.933 & 0.360 & 0.000 \\
\hline
\end{tabular*}\label{Tab:hk2}
\end{table}

\begin{table}
\caption{Relative entropy of clustering coefficient distribution.}
\begin{tabular*}{0.48\textwidth}{@{\extracolsep{\fill}}l | c c c c c c}
\hline
    & ER    & WS    & BA    & WG    & NLBA  & LSF   \\
\hline
 ER & 0.000 & 3.226 & 0.113 & 1.748 & 0.017 & 0.074 \\
 WS & 3.776 & 0.000 & 3.659 & 0.393 & 5.361 & 3.979 \\
 BA & 0.256 & 2.614 & 0.000 & 1.297 & 0.085 & 0.009 \\
 WG & 3.009 & 0.399 & 1.954 & 0.000 & 3.095 & 2.328 \\
 NLBA   & 0.029 & 3.052 & 0.056 & 1.616 & 0.000 & 0.028 \\
 LSF    & 0.147 & 2.738 & 0.008 & 1.385 & 0.037 & 0.000 \\
\hline
\end{tabular*}\label{Tab:cc}
\end{table}

\begin{table}
\caption{Relative entropy of hierarchical clustering coefficient of level 2 distribution.}
\begin{tabular*}{0.48\textwidth}{@{\extracolsep{\fill}}l | c c c c c c}
\hline
    & ER    & WS    & BA    & WG    & NLBA  & LSF   \\
\hline
 ER & 0.000 & 5.581 & 2.149 & 5.311 & 0.872 & 1.976 \\
 WS & 5.741 & 0.000 & 1.955 & 1.674 & 2.630 & 1.879 \\
 BA & 2.054 & 2.143 & 0.000 & 4.148 & 0.185 & 0.050 \\
 WG & 7.366 & 1.642 & 4.859 & 0.000 & 6.107 & 5.262 \\
 NLBA   & 1.196 & 2.738 & 0.249 & 4.446 & 0.000 & 0.227 \\
 LSF    & 2.371 & 1.794 & 0.052 & 3.991 & 0.200 & 0.000 \\
\hline
\end{tabular*}\label{Tab:hcc2}
\end{table}

\begin{table}
\caption{Relative entropy of hierarchical divergence ratio of level 2 distribution.}
\begin{tabular*}{0.48\textwidth}{@{\extracolsep{\fill}}l | c c c c c c}
\hline
    & ER    & WS    & BA    & WG    & NLBA  & LSF   \\
\hline
 ER & 0.000 & 2.701 & 0.363 & 2.922 & 0.118 & 0.280 \\
 WS & 4.849 & 0.000 & 1.948 & 0.588 & 2.197 & 1.785 \\
 BA & 0.484 & 1.564 & 0.000 & 2.514 & 0.130 & 0.027 \\
 WG & 2.880 & 0.766 & 3.269 & 0.000 & 3.110 & 2.602 \\
 NLBA   & 0.248 & 2.350 & 0.137 & 2.324 & 0.000 & 0.071 \\
 LSF    & 0.509 & 2.137 & 0.032 & 2.227 & 0.065 & 0.000 \\
\hline
\end{tabular*}\label{Tab:dr2}
\end{table}

\begin{table}
\caption{Relative entropy of shortest path length distribution.}
\begin{tabular*}{0.48\textwidth}{@{\extracolsep{\fill}}l | c c c c c c}
\hline
    & ER    & WS    & BA    & WG    & NLBA  & LSF   \\
\hline
 ER & 0.000 & 0.377 & 0.620 & 1.743 & 0.092 & 0.397 \\
 WS & 0.503 & 0.000 & 2.811 & 1.339 & 1.254 & 2.269 \\
 BA & 0.403 & 1.258 & 0.000 & 2.288 & 0.167 & 0.021 \\
 WG & 4.647 & 1.863 & 1.093 & 0.000 & 2.797 & 0.841 \\
 NLBA   & 0.068 & 0.683 & 0.214 & 1.952 & 0.000 & 0.096 \\
 LSF    & 0.266 & 1.064 & 0.022 & 2.184 & 0.079 & 0.000 \\
\hline
\end{tabular*}\label{Tab:l}
\end{table}

\begin{table}
\caption{Relative entropy of vertex betweenness centrality distribution.}
\begin{tabular*}{0.48\textwidth}{@{\extracolsep{\fill}}l | c c c c c c}
\hline
    & ER    & WS    & BA    & WG    & NLBA  & LSF   \\
\hline
 ER & 0.000 & 0.138 & 0.076 & 0.250 & 0.317 & 0.556 \\
 WS & 0.111 & 0.000 & 0.114 & 0.366 & 0.697 & 1.054 \\
 BA & 0.039 & 0.067 & 0.000 & 0.215 & 0.018 & 0.004 \\
 WG & 0.390 & 0.398 & 0.329 & 0.000 & 0.244 & 0.433 \\
 NLBA   & 0.286 & 0.656 & 0.025 & 0.204 & 0.000 & 0.046 \\
 LSF    & 0.435 & 0.892 & 0.010 & 0.342 & 0.053 & 0.000 \\
\hline
\end{tabular*}\label{Tab:bet}
\end{table}

\begin{table}
\caption{Average and standard deviation among the values present in Tables~\ref{Tab:degree} to~\ref{Tab:bet}.}
\begin{tabular}{ l|c }
\hline
Measurement & Av. and std \\
\hline
$hC_2$    &    2.69    (2.08)    \\
$hk_2$    &    1.92    (1.81)    \\
$C$    &    1.55    (1.57)    \\
$hdr_2$    &    1.51    (1.31)    \\
$k_{nn}$    &    1.19    (0.71)    \\
$\ell$    &    1.11    (1.10)    \\
$k$    &    0.56    (0.55)    \\
$B$    &    0.29    (0.27)    \\
\hline
\end{tabular}\label{Tab:av_disc}
\end{table}

\section{Conclusions}

Much of the success of complex network research has relied on the
accurate modeling of complex phenomena. To reach this goal, efforts
should be concentrated in developing methods able to obtain databases
and measurements that can characterize networks structures with
accuracy. Thus, the development of improved sampling techniques and
analysis of the behavior of measurements with respect to incomplete
networks or networks with biased connections are fundamental for
complex networks research. In this paper, we reported an analysis of
network measurements with respect to progressively perturbed
networks. The perturbations were performed at the edge level,
considering random removal, addition and rewiring. We applied the
relative entropy in order to quantify the robustness of the network
measurements considering six representative network models.  The four
measurements most suitable to analyze perturbed network were
identified as: the hierarchical clustering coefficient ($hC_2$),
hierarchical degree ($hk_2$), clustering coefficient ($C$) and
divergence ratio ($hdr_2$).  It is interesting to note that the node
degree did not result as the best network measurement, being
associated with poor discrimination between networks with distinct
structures. For instance, while the relative entropy between the ER
and WG model is just 0.005 when the degree distribution is considered,
it increases to 5.741 when the hierarchical clustering coefficient is
used instead. Among the network models, structures with scale-free
organization presented the highest robustness when submitted to
perturbations.  We suggest as future works the consideration of other
complex network measurements as well as other types of perturbations,
such as node removal or perturbation with preferential rules. The
consideration of multivariate statistical methods (e.g.\
MANOVA~\cite{Duda_Hart:01}) and data mining techniques can also help
complementing the perturbation and discrimination analysis.

\begin{acknowledgments}
  Luciano da F. Costa is grateful to FAPESP (05/00587-5), CNPq
  (301303/06-1) for financial support. Francisco A. Rodrigues
  acknowledges FAPESP sponsorship (07/50633-9), Paulino R. Villas Boas
  acknowledges CNPq sponsorship (141390/2004-2).
\end{acknowledgments}

\bibliographystyle{unsrt}
\bibliography{perturbation}

\end{document}